\documentclass[conference]{IEEEtran}
\IEEEoverridecommandlockouts

\usepackage{cite}
\usepackage{amsmath}
\usepackage{newtxtext,newtxmath} 
\usepackage{graphicx}
\usepackage{booktabs}
\usepackage{array}
\usepackage{url}
\usepackage{tikz}
\usetikzlibrary{arrows.meta,positioning}
\usepackage[hidelinks]{hyperref}
\usepackage{cleveref}

\begin{document}

\title{Toward the Right Analytical Model and System Software for Autonomous Driving Systems:\\ Open Problems and Research Directions}

\author{%
  \IEEEauthorblockN{Atsushi Yano\IEEEauthorrefmark{1}\IEEEauthorrefmark{2}
    and Takuya Azumi\IEEEauthorrefmark{3}\IEEEauthorrefmark{2}}
  \IEEEauthorblockA{\IEEEauthorrefmark{1}Graduate School of Science and Engineering,
    Saitama University, Japan}
  \IEEEauthorblockA{\IEEEauthorrefmark{2}TIER IV Incorporated, Japan}
  \IEEEauthorblockA{\IEEEauthorrefmark{3}Academic Association
    (Graduate School of Science and Engineering),
    Saitama University, Japan}
}

\maketitle

\begin{abstract}
  Autonomous driving (AD) systems continuously transform multi-rate and asynchronous sensor streams into vehicle actuation through graphs of callbacks, nodes, and middleware components.
  In such systems, temporal correctness cannot be characterized by the execution time or deadline of an individual task alone: localization and perception chains run in parallel, fuse data with different timestamps, converge at planning, and propagate through control to actuation.
  Moreover, the demand for high processing capability places AD systems on high-performance processors with multicore parallelism and GPU acceleration, where execution times vary strongly with the input scene, hardware state, and co-running work.
  Rare deadline misses at runtime therefore cannot be ruled out, and safety is preserved through fail-safe mechanisms such as the minimal-risk maneuver (MRM).
  This raises a two-sided question: what analytical models are needed to reason about timing in AD systems, and what system software is needed to realize, observe, and enforce those models on real platforms? On the analytical side, real-time research has evolved from periodic/sporadic tasks, directed acyclic graphs (DAGs), pipelines, mixed-criticality systems, and timer-/event-driven models toward end-to-end latency along cause-effect chains, data freshness, timing disparity, probabilistic timing, highest-criticality fail-safe operation, and early deadline-miss detection.
  On the system-software side, AD stacks and middleware, such as Autoware and ROS~2, expose both the opportunities and limitations of implementing analyzable timing behavior through executors, communication layers, tracing tools, and evaluation frameworks.
  This paper surveys these two lines of work and identifies the remaining gaps along five dimensions: units of timing constraints, timing metrics, resource models, execution-time variability, and safety integration.
  Rather than proposing a single new model or runtime, we formulate open problems and research directions for converging theory and practice: analytical models must move closer to AD reality, while AD system software must be reshaped into analyzable, enforceable, and safety-aware infrastructure.
\end{abstract}

\begin{IEEEkeywords}
  real-time systems, autonomous driving, analytical timing model, system
  software, ROS~2, Autoware, cause-effect chains, end-to-end latency, data
  freshness, probabilistic timing, deadline-miss detection, open problems.
\end{IEEEkeywords}

\section{Introduction}
\begin{figure*}[t]
  \centering
  \includegraphics[width=\textwidth]{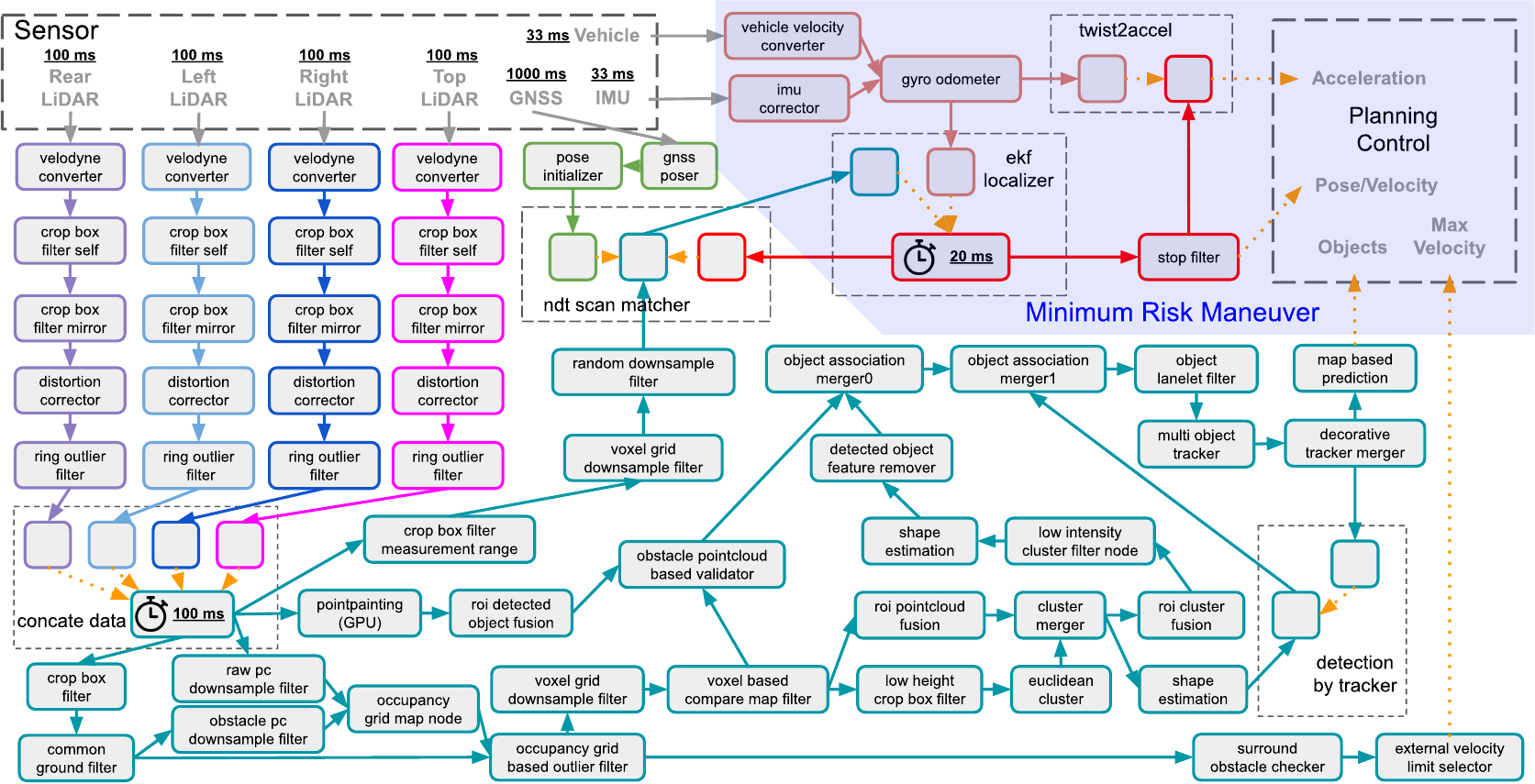}\\[4pt]
  \includegraphics[width=0.8\textwidth]{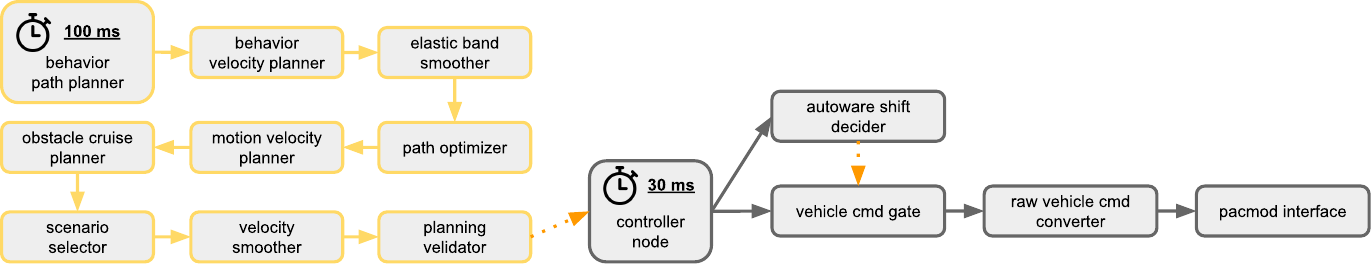}
  \caption{Main callbacks and their trigger mechanisms in Autoware~\cite{yano2024multideadline}.
    Sensor callbacks trigger two parallel chains: Localization (ndt scan matcher and ekf localizer with a 20\,ms deadline) and Perception (GPU-based pointpainting, object detection, and tracking driven by a 100\,ms timer).
    Both chains converge at Planning and proceed to Actuation via Control.
    The sensor pipeline required by ekf localizer for dead reckoning, ekf localizer itself, and parts of the subsequent localization, planning, and control are replicated for the minimal-risk maneuver (MRM)~\cite{kambe2025redundant}.}
  \label{fig:pipeline}
\end{figure*}

An autonomous driving (AD) system is a cyber-physical system that continuously transforms multiple sensor streams into vehicle actuation through a graph of callbacks rather than a single linear chain (\autoref{fig:pipeline}).
This figure shows one representative Autoware configuration, a concrete instance rather than a fixed architecture.
In it, each gray box is a Robot Operating System (ROS)~2 callback (or the node itself when the node holds a single callback), a dashed enclosure groups the callbacks of a multi-callback node, and solid edges carry publish/subscribe communication over a topic.
Orange dotted edges instead carry data through a node-internal queue or the ROS~2 take API~\cite{rclcpptake}; such an edge does not trigger the destination callback directly but affects its execution time and output.
Each callback is triggered by a timer at a fixed period (marked with a timer icon), by a subscription to a single topic, or by the synchronization of several topics (a sync callback, using the ApproximateTime or ExactTime policy~\cite{rosmessagefilters}).
Multi-rate sensors, such as multiple LiDARs, a global navigation satellite system (GNSS), and an inertial measurement unit (IMU), feed two parallel chains: a localization chain (ndt scan matcher and ekf localizer, constrained by a 20\,ms deadline) and a perception chain (GPU-based pointpainting, object detection, and tracking driven by a 100\,ms timer).
These chains converge at planning and propagate through control (a 30\,ms timer) to actuation.
In addition, some of these nodes, such as ekf localizer, planning, and control, are replicated for the minimal-risk maneuver (MRM), an emergency stop based on dead reckoning, and the replicas reside both on the same host and on a redundant electronic control unit (ECU)~\cite{kambe2025redundant}.
In such a graph, it is not sufficient that each node eventually produces a correct result.
The sensor data must not become too stale, the timestamps of data fused from different sensors must lie within a tolerable window, and a control command must be produced within a time that is still physically meaningful for the moving vehicle.
Timing correctness is therefore a property of the whole sensor-to-actuation graph rather than of any single task.

Classical real-time systems theory analyzes the schedulability of task sets built on periodic and sporadic tasks, worst-case execution time (WCET) and worst-case response time (WCRT), deadlines, and fixed-priority or earliest-deadline-first (EDF) scheduling~\cite{liu1973scheduling,audsley1993applying}.
In AD software, these assumptions are repeatedly strained.
(i)~The software is not a single task but many callbacks and nodes forming a graph.
(ii)~Sensor inputs and events are not co-periodic; the system is simultaneously multi-rate and event-driven.
(iii)~Because AD systems demand high processing capability, they run on modern processors with rich acceleration features, such as caches, speculative execution, multicore parallelism, and offloading to GPUs and neural processing units (NPUs).
On these platforms, CPU, GPU, NPU, communication, and memory contend in complex ways.
(iv)~Execution time varies strongly not only with the input scene and data volume but also with the hardware state and co-running work.
A single fixed WCET bound therefore becomes overly pessimistic, while a safety margin added to measured maxima can underestimate the true worst case.
(v)~Because absolute timing guarantees are impractical under such variability, rare deadline misses at runtime cannot be ruled out, and AD systems preserve safety against them through fail-safe mechanisms such as fallback, degraded modes, and the MRM~\cite{kambe2025redundant}.
A timing violation must therefore be interpreted not merely as a deadline miss but together with these safety mechanisms.

This paper asks what the right \emph{analytical model} and \emph{system software} for AD systems should be, and organizes the answer from two directions.
\autoref{sec:theory} surveys the analytical timing models and identifies five concrete gaps between classical assumptions and AD reality (\autoref{tab:gaps}).
\autoref{sec:impl} maps the implementation and evaluation research, centered on Autoware and ROS~2, that both makes those models observable and controllable on real systems and reveals where real systems break the assumptions theory relies on (\autoref{tab:landscape}).
\Cref{sec:openresearch,sec:openindustry} then pose the open problems that remain as two parallel lists: open problems for the real-time \emph{theory} community, and open problems for \emph{practitioners} who build and deploy AD software, respectively.
Our central argument is that this gap cannot be closed from either side alone: the theory must evolve toward AD reality, and AD implementations must in turn be reshaped so that they faithfully realize analyzable models.
Our contribution is therefore not a single new task model but (i)~a structured account of which existing models must be combined and where the theory--implementation gap is still open, and (ii)~a pair of open-problem agendas, one for each community, that together define this two-way convergence.

\section{Analytical Models and the Theory--Reality Gap}
\label{sec:theory}

\subsection{Timing Requirements in Real AD Systems}
Six timing requirements recur in AD systems, and a single deployed AD system must satisfy all of them at once.

\textbf{(R1)~End-to-end (E2E) latency along cause-effect chains:} the total delay from sensor input to actuator output must be bounded even when every individual node meets its deadline~\cite{becker2017e2e,teper2022e2e,gunzel2025phasing,betz2025containerized}.
Bounding this delay requires tracing which input sample influences which output decision as a sensor reading propagates through perception into planning and control~\cite{becker2017e2e,sun2023guaranteed,kobayashi2025rta}.

\textbf{(R2)~Response time and schedulability:} the classical question of whether a unit of work activated at a single rate, such as a sporadic or periodic task, a callback chain, or a directed acyclic graph (DAG), completes within its deadline, now posed on top of middleware executors~\cite{choi2021picas,sobhani2023timing,teper2025reconciling}.

\textbf{(R3)~Data freshness and timing disparity:} the age of data and the timestamp gap between fused sensor streams must be controlled; for example, LiDAR and camera results do not necessarily describe the environment at the same instant~\cite{li2022disparity,tang2024refreshing,sun2025disparity}.

\textbf{(R4)~Probabilistic timing:} because a fixed WCET is overly pessimistic while the average case is unsafe, timing should be expressed probabilistically, for example as the exceedance probability of the reaction time given input execution-time distributions~\cite{gunzel2023probabilistic}, or as probabilistic guarantees on E2E latency~\cite{han2023minimizing}.
Probabilistic WCET (pWCET) makes this concrete by upper-bounding the probability that the execution time exceeds a given value, so that a timing bound can be matched to the failure rate tolerated by the safety goal~\cite{gil2017openchallenges,bozhko2023pwcet}.

\textbf{(R5)~Highest-criticality runtime fail-safe:} because the rare misses that such probabilistic bounds tolerate can still occur at runtime, a fail-safe mechanism such as the MRM must be provisioned to preserve safety when they do.
As the last line of defense, it must remain operational precisely when the nominal functions are failing; therefore, its execution must be guaranteed at the highest criticality~\cite{burns2022mixed}.

\textbf{(R6)~Deadline-miss early detection:} the possibility of a miss should be detected \emph{during} execution, early enough to trigger such a safe transition rather than after the fact~\cite{toba2024deadline,toba2026enhancing}.

Recent frameworks make several of these metrics, notably maximum reaction time (MRT) and maximum data age (MDA), first-class, both in ROS~2~\cite{teper2022e2e,tang2024refreshing} and in CyberRT/Apollo~\cite{alcon2025cyberrt}.

\subsection{Existing Models and Their Assumptions}
None of these requirements is uncharted: each is attacked by an active line of work built from a small set of building blocks.
For E2E latency~(R1), cause-effect-chain analysis bounds reaction time and data age over ROS~2 callback graphs~\cite{teper2022e2e}, and E2E latency is measured and optimized for containerized ROS~2 AD software~\cite{betz2025containerized}.
For response time~(R2), the periodic/sporadic task model and its response-time analysis are re-instantiated on middleware executors via chain-aware priorities~\cite{choi2021picas}, multi-threaded executor analysis~\cite{sobhani2023timing}, and executor designs that restore the applicability of classical scheduling theory~\cite{teper2025reconciling}.
DAG and precedence models expose intra-task parallelism and dependencies~\cite{sun2023guaranteed,kobayashi2025rta}, and timer-driven/event-driven activation models capture how nodes are released~\cite{teper2022e2e,sun2025disparity}.
For freshness and disparity~(R3), synchronization and buffering policies are modeled to bound worst-case time disparity~\cite{li2022disparity}, data age under refreshing~\cite{tang2024refreshing}, and disparity jointly with E2E latency~\cite{sun2025disparity}.
For probabilistic timing~(R4), pWCET frameworks~\cite{gil2017openchallenges,bozhko2023pwcet} and probabilistic analyses of reaction time~\cite{gunzel2023probabilistic} and E2E latency~\cite{han2023minimizing} replace single worst-case bounds with exceedance probabilities.
For the highest-criticality fail-safe~(R5), mixed-criticality scheduling co-hosts functions of different safety importance and preserves the guarantees of the most critical ones when optimistic execution-time budgets are exceeded~\cite{burns2022mixed}.
For deadline-miss early detection~(R6), runtime checks of DAG progress against intermediate thresholds detect an impending miss during execution~\cite{toba2024deadline,toba2026enhancing}.

These models are indispensable, but each line of work addresses its requirement largely in isolation, and how to combine them into one analysis that covers all six requirements for the same system remains open.
Moreover, they are usually instantiated under assumptions that hold poorly for AD software: WCET is fairly well known, and computing resources other than the CPU are either not modeled or abstracted away.

\subsection{Five Gaps}
These assumptions diverge from AD reality along five dimensions, summarized in \autoref{tab:gaps}.

\textbf{(G1)~Unit of timing constraint:} classical analysis imposes the deadline on the same clear unit task that the scheduler dispatches, whereas in ROS~2/Autoware the schedulable unit is the callback while safety-relevant constraints such as E2E latency attach to an E2E cause-effect chain or the whole graph.
Which unit (node, callback, chain, E2E cause-effect chain, or graph) should carry a given timing constraint is therefore not obvious.

\textbf{(G2)~Metric integration:} deadlines alone are insufficient when age, freshness, jitter, disparity, and E2E latency matter at once.

\textbf{(G3)~Resource modeling:} CPU-only models cannot express GPU/NPU inference, memory bandwidth, and many-core communication~\cite{azumi2020roslite}.

\textbf{(G4)~Execution-time variability:} execution time depends strongly on the number of objects, point-cloud density, and scene complexity; thus, a single worst case or average is rarely actionable.
Contention for shared resources on multicore and heterogeneous systems-on-chip (SoCs) adds hardware-induced variability~\cite{reghenzani2020uncertainty,nowotsch2014interference,brilli2022cmri,bateni2020coopt}, motivating pWCET timing models~\cite{gil2017openchallenges,bozhko2023pwcet}.

\textbf{(G5)~Safety integration:} safety assurance is meaningful only in connection with redundancy, degraded modes, and the MRM.
It must therefore extend to runtime: hazardous states such as an impending deadline miss must be detected during execution, and the fail-safe mechanism triggered in response must be guaranteed to execute.

\begin{table*}[t]
  \caption{Where classical real-time assumptions diverge from AD reality.}
  \label{tab:gaps}
  \centering
  \footnotesize
  \renewcommand{\arraystretch}{1.25}
  \begin{tabular}{@{}p{0.22\textwidth} p{0.355\textwidth} p{0.325\textwidth}@{}}
    \toprule
    \textbf{Aspect}            & \textbf{Classical real-time model}       & \textbf{AD reality}                                                                   \\
    \midrule
    Unit of timing constraint  & deadline on a clear unit task            & constraints at chain/graph level, scheduling at callback level; units do not coincide \\
    Timing metric              & deadline-centric (period, response time) & age, freshness, jitter, disparity, E2E latency                                        \\
    Resource                   & CPU-centric                              & CPU/GPU/NPU, communication, memory, IPC contention                                    \\
    Execution-time variability & small / bounded                          & scene-/data-dependent \emph{and} hardware-induced (shared-resource contention), large \\
    Safety                     & external, post-hoc                       & redundancy, degraded mode, MRM; runtime hazard detection, fail-safe execution         \\
    \bottomrule
  \end{tabular}
\end{table*}

\subsection{Requirements of a Practical Real-Time Task and Timing Model}
The practical model is therefore not one new task model but a layered combination of the existing building blocks.
This combination (a)~takes the callback chain/graph, not a single task, as the unit of analysis; (b)~integrates multi-rate and event-driven activation; (c)~evaluates E2E latency together with freshness, jitter, and timing disparity; (d)~incorporates heterogeneous CPU/GPU/NPU, communication, and memory contention; (e)~uses variability-aware execution-time models that reflect artificial intelligence (AI) workloads, scene changes, and platform-level resource contention; and (f)~connects to safety control via criticality, fallback, and the MRM, including runtime detection of impending deadline misses and guaranteed execution of the triggered fail-safe mechanism.

\section{Implementation and Evaluation Landscape}
\label{sec:impl}
The gap of \autoref{sec:theory} is not only that theory abstracts away AD reality; the implementation side contributes its own half, since real AD stacks are not built to be analyzed.
\autoref{tab:landscape} organizes the implementation and evaluation research along five layers, summarized below.

\begin{table*}[t]
  \caption{Implementation and evaluation landscape for AD real-time systems, centered on Autoware and ROS~2.}
  \label{tab:landscape}
  \centering
  \footnotesize
  \renewcommand{\arraystretch}{1.3}
  \begin{tabular}{@{}p{0.16\textwidth} p{0.52\textwidth} p{0.25\textwidth}@{}}
    \toprule
    \textbf{Layer}                                                                                                                                                                                  & \textbf{Representative artifacts} & \textbf{Timing aspect addressed} \\
    \midrule
    Application                                                                                                                                                                                     &
    Autoware~\cite{kato2018autoware}                                                                                                                                                                &
    realistic chains / nodes / topics                                                                                                                                                                                                                                      \\
    Executor                                                                                                                                                                                        &
    CallbackGroup executor~\cite{yang2020exploring}, ThreadedCallback~\cite{peng2021threaded},
    PiCAS~\cite{choi2021picas}, multi-threaded executor timing~\cite{sobhani2023timing},
    classical real-time reconciliation~\cite{teper2025reconciling}, ROSRT~\cite{liu2025rosrt},
    callback enforcement~\cite{ishikawa2025callback}, and ROSCH~\cite{saito2018rosch}                                                                                                               &
    callback dispatch, chain priority, and response time                                                                                                                                                                                                                   \\
    Communication                                                                                                                                                                                   &
    ROS~2/DDS transport evaluation~\cite{maruyama2016exploring}, NoC communication~\cite{azumi2020roslite, tajima2024roslite2},
    TZC~\cite{wang2019tzc}, iceoryx~\cite{poehnl2023iceoryx},
    Agnocast~\cite{ishikawa2025agnocast}, and CROS-RT~\cite{kim2025crosrt}                                                                                                                          &
    IPC / zero-copy, NoC, memory movement, and cross-layer priority                                                                                                                                                                                                        \\
    Tracing \& evaluation                                                                                                                                                                           &
    ros2\_tracing~\cite{bedard2022ros2tracing}, CARET~\cite{kuboichi2022caret},
    TILDE~\cite{he2023tilde}, Autoware\_Perf~\cite{li2022autowareperf}, and RD-Gen~\cite{yano2023rdgen}                                                                                             &
    latency tracing, freshness / deadline-miss detection, and reproducible multi-rate DAG benchmarks                                                                                                                                                                       \\
    OS \& hardware                                                                                                                                                                                  &
    Autoware on Board~\cite{kato2018autoware},
    ROS-lite~\cite{azumi2020roslite} and ROS-lite2~\cite{tajima2024roslite2}, heterogeneous SoC~\cite{chishiro2019heterogeneous}, and virtualized / containerized execution~\cite{wen2023baremetal} &
    OS scheduling, affinity, bandwidth, and virtualization overhead                                                                                                                                                                                                        \\
    \bottomrule
  \end{tabular}
\end{table*}

\emph{Application.} Autoware is an open-source AD stack where abstract DAGs, pipelines, and cause-effect chains materialize as concrete callbacks, nodes, topics, and executors~\cite{kato2018autoware}.
It is increasingly the common benchmark for AD timing studies~\cite{becker2020demystifying,peeck2021monitoring,betz2023latency,teper2023timingaware,abaza2024trace,sciangula2024dds,fan2026uncovering}.

\emph{Executor.} In ROS~2, the executor connects a scheduling policy to actual callback dispatch.
Existing work restructures the dispatch mechanism itself (CallbackGroup~\cite{yang2020exploring}, thread-per-callback~\cite{peng2021threaded}, and one-to-one callback-to-thread enforcement~\cite{ishikawa2025callback}), imposes scheduling policies on it (ROSCH~\cite{saito2018rosch}, chain-aware priorities~\cite{choi2021picas}, and flexible OS-level scheduling~\cite{liu2025rosrt}), or analyzes the resulting timing (multi-threaded executors~\cite{sobhani2023timing} and reconciliation with classical periodic-task scheduling~\cite{teper2025reconciling}).
Together they show that the executor is itself part of the timing model rather than a transparent layer the analysis can ignore.

\emph{Communication.} Data movement matters as much as scheduling: an early ROS~2 study quantified Data Distribution Service (DDS) transport latency, throughput, and memory overhead across vendors and quality-of-service (QoS) settings~\cite{maruyama2016exploring}, and many-core platforms move inter-node communication onto a network-on-chip (NoC)~\cite{azumi2020roslite,tajima2024roslite2}.
A shared-memory and zero-copy line of work (TZC~\cite{wang2019tzc}, iceoryx~\cite{poehnl2023iceoryx}, and Agnocast~\cite{ishikawa2025agnocast}) removes serialization and copy costs from inter-process communication (IPC), while CROS-RT aligns priorities across the application, middleware, and kernel layers to bound IPC response times~\cite{kim2025crosrt}.

\emph{Tracing \& evaluation.} To observe timing on real systems, ros2\_tracing~\cite{bedard2022ros2tracing} (extended with message-flow analysis~\cite{bedard2023messageflow}), CARET~\cite{kuboichi2022caret}, and Autoware\_Perf~\cite{li2022autowareperf} measure E2E and path-level latency, while TILDE additionally tracks per-message data freshness and detects deadline misses at runtime~\cite{he2023tilde}.
For controlled experiments, RD-Gen generates reproducible multi-rate DAG workloads~\cite{yano2023rdgen}.

\emph{OS \& hardware.} Finally, behavior is shaped by the platform: Autoware on Board runs a full AD stack on Linux-based embedded GPU devices~\cite{kato2018autoware}, ROS-lite and ROS-lite2 port the ROS execution model onto real-time operating system (RTOS)-based many-core platforms~\cite{azumi2020roslite,tajima2024roslite2}, and heterogeneous SoCs host such stacks under power and real-time constraints~\cite{chishiro2019heterogeneous}.
Toward software-defined vehicles, hypervisor- and container-based deployment of Autoware achieves near-bare-metal performance except for disk I/O~\cite{wen2023baremetal}.

Each surveyed effort strengthens timing at a single layer, but no combination of them yet yields a stack whose whole-system safety can be fully guaranteed.
Taken together, \Cref{sec:theory,sec:impl} show that the theory--reality gap has two facing edges: models that abstract away AD reality, and implementations that remain hard to analyze.
Closing the gap is therefore a problem of \emph{mutual convergence}: theory must move toward AD reality while implementations move toward analyzable structure, and neither community can close it alone.
We make this concrete as two parallel sets of open problems: \autoref{sec:openresearch} asks what \emph{model} the theory community still owes, and \autoref{sec:openindustry} asks what \emph{system} practitioners must build to meet it.

\section{Open Problems: For Theorists}
\label{sec:openresearch}
The first set of open problems targets the real-time \emph{theory} community.
Building on the five gaps of \autoref{tab:gaps} and the requirements of \autoref{sec:theory}, each problem below asks what new task or timing \emph{model}, rather than what new implementation, is still missing.

\textbf{(TP1)~The right unit of analysis:} What should be the unit of analysis: callback, node, chain, E2E cause-effect chain, or graph?
Classical schedulability assumes a clear unit task, but in ROS~2 the schedulable unit (the callback) and the safety-relevant unit (the cause-effect chain or graph) do not coincide~\cite{choi2021picas,teper2025reconciling}.
Chain- and DAG-level models exist~\cite{sun2023guaranteed,yano2024multideadline}, yet a single model whose unit simultaneously carries schedulability and safety meaning is still open.
Theorists and practitioners need to converge, through dialogue, on a unit that is flexible enough to express the implementation structures that AD systems essentially require, yet faithful to the timing-constraint demands of the real system.

\textbf{(TP2)~Joint analysis of heterogeneous timing metrics:} How can one analysis reason jointly about E2E latency, response time, freshness, timing disparity, and miss probability?
These metrics carry different safety meanings, and only a few pairs are analyzed together today: disparity with E2E latency~\cite{li2022disparity,sun2025disparity} and MRT with MDA~\cite{tang2024refreshing,alcon2025cyberrt}.
Collapsing them into one number loses information; an analytical model that treats the full set jointly remains open.
One direction is a vector-valued timing contract that constrains the metrics jointly, e.g., bounding freshness and disparity while optimizing E2E latency, so that their interactions are analyzed rather than averaged away.

\textbf{(TP3)~Heterogeneous resource models:} How can a timing model explicitly capture GPU/NPU execution, memory bandwidth, and on-chip interconnect, rather than leaving them unmodeled or abstracted away?
NoC-based and clustered many-core ports of the ROS execution model~\cite{azumi2020roslite,tajima2024roslite2}, and heterogeneous SoCs~\cite{chishiro2019heterogeneous} show that these resources are inside the timing budget; an analysis of ROS~2 systems that makes them first-class is open.

\textbf{(TP4)~Execution-time models under scene-/data-dependent and hardware-induced variability:} How should execution time be modeled when it depends on scene, data volume, and deep neural network (DNN) pre-/post-processing?
A single worst case or average is rarely actionable, and scene-dependent workload should be separated from hardware- and runtime-induced residual variation.
The former is largely fixed once the scene is given (e.g., the workload scales with the number of detected objects or input points); thus, subsuming it into a probability distribution mischaracterizes it.
Probabilistic timing~\cite{gunzel2023probabilistic,han2023minimizing} is a starting point; a model that captures the scene-dependent workload as an explicit function of the scene and only the residual variation probabilistically is open.

\textbf{(TP5)~Binding schedulability to safety across the MRM mode change:} What does a timing violation \emph{mean} for safety, and what must a timing analysis guarantee when a miss becomes imminent?
A deadline miss or stale datum acquires meaning only once tied to fallback, degraded modes, and the MRM; therefore, the analysis should cover the full fail-safe sequence: detecting an impending miss at the optimal instant (too late leaves no time for a safe reaction; too early causes spurious mode changes), triggering the mode change that starts the MRM, and guaranteeing MRM schedulability after the change.
Schedulability analysis~\cite{sun2023guaranteed}, runtime miss detection~\cite{toba2024deadline,toba2026enhancing}, and mixed-criticality scheduling~\cite{burns2022mixed} each address one stage in isolation; a model that co-derives all three remains open.

\section{Open Problems: For Practitioners}
\label{sec:openindustry}

The second set of open problems targets \emph{practitioners} who build and deploy AD software on ROS~2 and Autoware.
Where \autoref{sec:openresearch} asks what model is missing, each problem below asks how to \emph{engineer} a real system that is at once analyzable, performant, and safe.
Each is either the practitioner-side dual of a research problem or a necessary condition for applying the analyses those problems call for to a real system.
As duals, heterogeneous resource modeling (TP3) meets resource isolation and middleware transparency (PP1,~PP3), and binding schedulability to safety across the MRM mode change (TP5) meets online violation detection (PP4).
As necessary conditions, no analysis applies unless the implementation faithfully realizes the task model (PP2) and the platform enforces the scheduler that the analysis assumes (PP5).
Progress on one side is therefore useful only if matched on the other.

\textbf{(PP1)~Eliminating non-essential complexity:} How can we remove the non-essential complexity that the middleware layer imposes on scheduling?
In ROS~2, nested scheduling (the ROS~2 executor on top of the OS thread scheduler) is an implementation artifact rather than an inherent necessity.
Establishing a persistent one-to-one correspondence between callbacks and OS threads lets OS scheduling parameters apply directly per callback and lets analysis ignore the executor altogether~\cite{ishikawa2025callback}.
Communication is a second source: DDS runs its own internal threads for message transport, invisible to the scheduling model yet contending for the same cores; adopting true zero-copy IPC such as Agnocast eliminates these dedicated DDS communication threads~\cite{ishikawa2025agnocast}.
Even so, residues remain: a mutually exclusive CallbackGroup serializes its callbacks outside the view of the OS scheduler, an overrun of a non-reentrant callback reintroduces middleware-level queueing, and discovery, services, and inter-host transport still occupy DDS threads.
Eliminating these residues and characterizing exactly when the middleware layer can be ignored is open.

\textbf{(PP2)~Enforcing the implementation--model contract:} How can the implementation be shaped so that it faithfully realizes the theoretical task model?
Task models presume that a task makes its outputs visible to successors only upon completion, yet the ROS~2 publish/subscribe API does not force the message sending of a task to occur at its end: a \texttt{publish} may fire at any point during execution.
This unconstrained send timing breaks not only DAG precedence (completion boundaries and join synchronization) but the precedence assumptions of task models in general; hence, the model holds only by programmer discipline and collapses once violated.
Expressing each subtask as a function whose arguments and return values are its incoming and outgoing edges binds message exchange to subtask completion at the API rather than by convention, opening a direct path from scheduling theory to practice~\cite{ishikawa2025fass}.
Extending this to message synchronization, shared-variable, and blocking flows and deploying it on production middleware is open.

\textbf{(PP3)~Separating functions of different criticality:} How can functions of different criticality (the safety-critical MRM at the highest level, the nominal AD functions of perception, planning, and control at an intermediate level, and logging or a human-machine interface at the lowest) share one platform without the lower-criticality ones disturbing the higher-criticality ones?
The underlying tension is between \emph{partitioning} for assurance and \emph{sharing} for efficient resource usage~\cite{burns2022mixed}: full physical separation isolates criticalities but squanders the limited compute, power, and weight budget of an embedded platform; the goal is therefore to consolidate functions on shared hardware while still guaranteeing isolation.
Commodity ROS~2 on Linux offers only partial mechanisms: running real-time DAGs under FIFO/EDF while confining best-effort work to the Completely Fair Scheduler (CFS) on exclusive CPU affinities~\cite{ishikawa2025callback}, hosting on heterogeneous SoCs~\cite{chishiro2019heterogeneous}, and virtualized or containerized deployment~\cite{betz2025containerized,wen2023baremetal}.
Enforcing temporal and spatial isolation by criticality (without statically over-provisioning away the efficiency that motivated sharing) and proving it holds on such commodity stacks is open.

\textbf{(PP4)~Detecting timing-constraint violations early:} How can we detect, at runtime and early enough to act, that a timing constraint will be missed?
Post-hoc detection is too late for a safe transition.
Early-detection methods for DAG tasks with variable and probabilistic thresholds~\cite{toba2024deadline,toba2026enhancing} and runtime miss/freshness tracking in tools such as TILDE~\cite{he2023tilde} point the way.
A low-overhead, chain- and graph-level online detector wired to fallback and deployable in production remains open.

\textbf{(PP5)~Implementing state-of-the-art schedulers:} How can the latest scheduling algorithms actually be enforced on commodity middleware and OS?
The underlying obstacle is that no system software treats a timing-constrained unit of execution (a DAG, for instance) as a \emph{first-class citizen}: an entity the runtime maintains natively and whose precedence and deadline semantics the scheduler enforces directly, rather than one reconstructed by convention atop primitives (threads or callbacks) that have no notion of it.
Lacking such a substrate, published DAG schedulers cannot be enforced as specified and are evaluated largely by analysis or simulation.
Enabling mechanisms are emerging: per-callback OS parameters~\cite{ishikawa2025callback}, chain-aware priorities~\cite{choi2021picas}, flexible OS-level scheduling~\cite{liu2025rosrt}, and reconciliation with classical periodic scheduling~\cite{teper2025reconciling}.
None of them, however, makes the timing-constrained unit first-class.
A repeatable path from a published algorithm to a low-overhead, enforced implementation on production stacks is open.

\section{Conclusion}
\label{sec:conclusion}
This paper revisited real-time models for AD systems from both the theoretical and the implementation/evaluation side.
On the theory side, work built on periodic/sporadic tasks, DAGs, pipelines, mixed-criticality, and timer-/event-driven activation is evolving toward E2E latency along cause-effect chains, data freshness, timing disparity, probabilistic timing, highest-criticality fail-safe operation, and early miss detection.
On the implementation side, Autoware, ROS~2 executors, ROS-lite and ROS-lite2, Agnocast, and tracing/evaluation tools such as CARET, TILDE, ros2\_tracing, and Autoware\_Perf measure and improve real-time behavior on real systems.
A practical real-time task and timing model for AD systems is therefore not a single task model but a layered framework spanning callback/graph/chain units, multi-rate/event-driven execution, chain-aware metrics, heterogeneous resources, variability-aware timing, and a connection to safety control.
Closing the theory--reality gap is a two-way convergence: theoretical models must evolve toward AD reality, and Autoware/ROS~2 implementations must be reshaped to realize analyzable models.
The key next step is to make both sides move by binding these models to Autoware/ROS~2 tooling while reshaping that tooling to be analyzable, so that theory can inform practical design and practice can ground theory.
We framed these next steps as two parallel sets of open problems: for the theory community (the right unit of analysis, joint metrics, heterogeneous resources, execution-time variability, and the safety connection) and for practitioners (removing non-essential complexity, closing the implementation--model gap, isolating functions of different criticality, online violation detection, and deploying state-of-the-art schedulers).

\section*{Acknowledgments}
This work was supported by JST FOREST Grant Number JPMJFR242G, and is based on results obtained from a project, Green Innovation Fund Projects / Development of In-vehicle Computing and Simulation Technology for Energy Saving in Electric Vehicles (JPNP21027), subsidized by the New Energy and Industrial Technology Development Organization (NEDO).

\bibliographystyle{IEEEtran}
\bibliography{reference}

\end{document}